\documentclass[a4paper, 12pt]{article}
\usepackage{a4wide}
\usepackage[normalem]{ulem}
\usepackage{dsfont}
\usepackage[T1]{fontenc}
\usepackage{amsmath}
\usepackage{amssymb}
\usepackage{amsthm}
\usepackage[ansinew]{inputenc}
\usepackage{enumerate}
\usepackage{fancyhdr}
\usepackage{cite}
\usepackage{color}
\usepackage{ifpdf}
\usepackage{graphicx}
\usepackage{rotating}
\newtheorem{theorem}{Theorem}

\DeclareMathOperator{\tr}{tr}

\newcommand{\R}        {\mathds{R}}

\begin{document}
\title{Isotropization of non-diagonal Bianchi I spacetimes with collisionless matter at late times assuming small data}
\author{Ernesto Nungesser\\ 
Max-Planck-Institut f\"ur Gravitationsphysik\\
Albert-Einstein-Institut\\Am M\"uhlenberg 1\\
14476 Potsdam, Germany\\
ernesto.nungesser@aei.mpg.de}
\maketitle

\begin{abstract}
Assuming that the space-time is close to isotropic in the sense that
the shear parameter is small and that the
maximal velocity of the particles is bounded, we have been able to show that for
non-diagonal Bianchi I-symmetric spacetimes with collisionless matter
the asymptotic behaviour at late times is close to the special case of
dust. We also have been able to show that all the Kasner exponents converge to
$\frac{1}{3}$ and an asymptotic expression for the induced metric
has been obtained. The key was a bootstrap argument.
\end{abstract}

The sign conventions of \cite{RA} are used. In particular, we use metric signature -- + + + and geometrized units, i.e. the gravitational constant G and the speed of light c are set equal to one. Also the Einstein summation convention that repeated indices are to be summed over is used. Latin indices run from one to three. $C$ will be an arbitrary constant and $\epsilon$ will denote a small and strictly positive constant. They both may appear several times in different equations or inequalities without being the same constant. A dot above a letter will denote a derivative with respect to the cosmological (Gaussian) time $t$.

\section{Introduction}

For both mathematical and physical reasons one would like to be able to understand anisotropic and inhomogeneous cosmological models in general. As a first step spatially homogeneous spacetimes can be studied. For the case where the matter model is a perfect fluid a lot of results have been already obtained which are summarized in \cite{WE}, see also \cite{HU} for a critical discussion. Usually in observational cosmology it is assumed that there exists an Era which is ``matter-dominated'' where the matter model is a perfect fluid with zero pressure, i.e. the dust model. A kinetic description via collisionless matter enables to study the stability of this model in the following sense. Suppose we have an expanding universe where the particles have certain velocity dispersion. One might think that due to the expansion the velocity dispersion will decay. That this is true has been shown in \cite{RT}, \cite{RU} and \cite{RE} for locally rotationally symmetric (LRS) models in the cases of Bianchi I, II and III. These results have been generalized (in a different direction) to LRS Bianchi IX and in a context which also goes beyond collisionless matter in \cite{PP} and \cite{CH}.\\The LRS models can be diagonalized in a suitable frame where they then stay diagonal if one makes some extra assumptions on the distribution function. A natural question is, what happens if the model is not diagonal? In this paper we will treat the late time dynamics of the \textit{non-diagonal} Bianchi I case assuming small data. That this makes sense was established in \cite{CC} where it was shown that geodesic completeness holds for the general Bianchi I-Vlasov case. In a sense which will be specified later, we assume that the universe is close to isotropic and that the velocity dispersion of the particles is bounded. Then we conclude that the universe will isotropize and have a dust-like behaviour asymptotically. In fact these two properties are intimately linked in the proof, something which is not expected to happen in other Bianchi types, except in the case of a positive cosmological constant where it has been shown \cite{Lee} that isotropization and asymptotic dust-like behaviour occurs in all Bianchi models except Bianchi IX and without the LRS assumption.\\ We hope to be able to extend this result in the near future a) to other Bianchi types, where the spacetime does not (necessarily) isotropize, but some other solution still may act as an 'attractor' b) to remove the small data assumption(s).\\

\section{Bianchi I spacetimes with collisionless matter}

A \textit{Bianchi spacetime} is defined to be a spatially homogeneous spacetime whose isometry group possesses a three-dimensional subgroup $G$ that acts simply transitively on the spacelike orbits. They can be classified by the structure constants of the Lie algebra associated to the Lie group. We will only consider the simplest case where the structure constants vanish, i.e. the case of Bianchi I with the abelian group of translations in $\R^3$ as the Lie group, where the metric has the following form using Gauss coordinates:
\begin{eqnarray}\label{me}
 ^4 g=-dt^2+g_{ab}(t)dx^adx^b.
\end{eqnarray}

We will use the 3+1 decomposition of the Einstein equations as made in \cite{RA}. We use Gauss coordinates, which implies that the lapse function is the identity and the shift vector vanishes, so comparing our metric with (2.28) of \cite{RA} we have that $\alpha=1$ and $\beta^a=0$. The only non-trivial Christoffel symbols are the following (See (2.44)-(2.49) of \cite{RA}):
\begin{eqnarray}
 \Gamma^0_{ab}&=&-k_{ab}\\
\Gamma^a_{0b}&=&-k^a_b
\end{eqnarray}
In terms of coordinate expressions we have then that ($n_0=-1$)
\begin{eqnarray*}
   \rho&=&T^{00}\\
     j_a&=&T_a^0\\
  S_{ab}&=&T_{ab}
\end{eqnarray*}
where $\rho$, $j_a$ and $T_{\mu\nu}$ are the energy density, matter current and energy momentum tensor respectively.
In the 3+1 formulation the second fundamental form $k_{ab}$ is used and rewriting its definition as in (2.29) of \cite{RA} we have:
\begin{eqnarray}\label{a}
 \dot{g}_{ab}=-2 k_{ab}.
\end{eqnarray}
Using the Einstein equations as in (2.34) of \cite{RA} and the fact that $g_{ab}$ does not depend on the spatial variables, we have:
\begin{eqnarray}\label{b}
 \dot{k}_{ab}=H k_{ab}-2k_{ac}k^c_b-8\pi(S_{ab}-\frac{1}{2}g_{ab}\tr S)-4\pi\rho g_{ab}
\end{eqnarray}
where we have used the notations $\tr S =g^{ab}S_{ab}$, $H=\tr k$. It has been assumed that the cosmological constant vanishes.
From the constraint equations (2.26)-(2.27) of \cite{RA}:
\begin{eqnarray}\label{c}
 -k_{ab}k^{ab}+H^2=16\pi\rho
\end{eqnarray}
and we see that no matter current is present since the ``spatial'' Christoffel symbols vanish:
\begin{eqnarray}
 T_{0a}=0.
\end{eqnarray}
From (\ref{c}) we see that $H$ never vanishes except for the Minkowski spacetime. We exclude this case and assume now without loss of generality that $H <0$ for all time. If this is not true for a given solution it may be arranged by doing the transformation $t\mapsto-t$.\\

For the matter model we will take the point of view of kinetic theory. This means that we have a collection of particles (in a cosmological context the particles are galaxies or clusters of galaxies) which are described statistically by a non-negative distribution function $f(x^\alpha,p^\alpha)$ which is the density of particles at a given spacetime point with given four-momentum. We will assume that all the particles have equal mass (one can relax this condition if necessary, see \cite{PP}). We want that our matter model is compatible with our symmetry assumption, so we will also assume that $f$ does not depend on $x^a$. In addition to that we will assume that there are no collisions between the particles. In this case the distribution function satisfies the Vlasov equation (See (3.38) of \cite{RA}):
\begin{eqnarray}
\frac{\partial f}{\partial t}+2 k^a_b p^b\frac{\partial f}{\partial p^a}=0.
\end{eqnarray}
where f is defined on the set determined by the equation 
\begin{eqnarray*}
 -(p^0)^2+g_{ab}p^ap^b=-m^2
\end{eqnarray*}
called the mass shell.
The energy momentum tensor is (compare with (3.37) of \cite{RA}):
\begin{eqnarray}
&& \rho=\int f(t,p)(m^2+g_{ab}p^ap^b)^{\frac{1}{2}}(\det g)^{\frac{1}{2}}dp\\
&&S_{ab}=\int f(t,p)p_ap_b(m^2+g_{ab}p^ap^b)^{-\frac{1}{2}}(\det g)^{\frac{1}{2}}dp\\\label{cu}
&&T_{0a}=\int f(t,p)p_a(\det g)^{\frac{1}{2}}dp
\end{eqnarray}
Here $p:=(p^1,p^2,p^3)$ and $dp:=dp^1 dp^2 dp^3$.
For this kind of matter all the energy conditions hold. In particular $\rho \geq \tr S \geq 0$.
Our system of equations consists of the equations (\ref{a})-(\ref{cu}). For a given Bianchi I geometry the Vlasov equation can be solved explicitly with the result that if $f$ is expressed in terms of the covariant components $p_i$ then it is independent of time. This has the consequence that if $t_0$ is some fixed time and $f_0(p_i)=f(t,p_i)$ then we can express the non-trivial components of the energy momentum tensor as follows:
\begin{eqnarray*}
&& \rho=\int f_{0}(p_{i})(m^2+g^{cd}p_{c}p_d)^{\frac{1}{2}}(\det g)^{-\frac{1}{2}}dp_{1}dp_{2}dp_{3}\\
&& S_{ab}=\int f_{0}(p_{i})p_a p_b(m^2+g^{cd}p_{c}p_d)^{-\frac{1}{2}}(\det g)^{-\frac{1}{2}}dp_{1}dp_{2}dp_{3}.
\end{eqnarray*}

\subsection{Other equations and new variables}

A useful relation concerns the determinant of the metric ((2.30) of \cite{RA}):
\begin{eqnarray}\label{det}
\frac{d}{dt}[\log (\det g)]=-2 H 
\end{eqnarray}
Taking the trace of the mixed version of the second fundamental form (2.36 of \cite{RA}):
\begin{eqnarray}\label{im}
 \dot{H}=H^2+4\pi \tr S -12\pi \rho
\end{eqnarray}
With (\ref{c}) one can eliminate the energy density and (\ref{im}) reads:
\begin{eqnarray}\label{in}
 \dot{H}=\frac{1}{4}(H^2+3k_{ab}k^{ab})+4\pi \tr S
\end{eqnarray}
 We can decompose the second fundamental form introducing $\sigma_{ab}$ as the trace-free part:
\begin{eqnarray}\label{TF}
 k_{ab}=\sigma_{ab}+\frac{H}{3}g_{ab}
\end{eqnarray}
Then
\begin{eqnarray}\label{tf}
 k_{ab}k^{ab}=\sigma_{ab}\sigma^{ab}+\frac{H^2}{3}
\end{eqnarray}
and (\ref{in}) takes the following form:
\begin{eqnarray}\label{yes}
\dot{H}=\frac{1}{2}H^{2}+\frac{3}{4}\sigma_{ab}\sigma^{ab}+4\pi \tr S
\end{eqnarray}
It is obvious that
\begin{eqnarray}\label{ie}
 \dot{H}\geq\frac{1}{2}H^{2}
\end{eqnarray}
From the constraint equation (\ref{c}) and (\ref{yes}) it also follows that
\begin{eqnarray}\label{up}
\dot{H}\leq H^2
\end{eqnarray}

From (\ref{ie}) and (\ref{up}) we conclude that
\begin{eqnarray}\label{if}
 -(t+C)^{-1}\geq H(t)\geq -2(t+C)^{-1}.
\end{eqnarray}

Let us use the following notation:
\begin{eqnarray}\label{AN}
 F=\frac{\sigma_{ab}\sigma^{ab}}{H^2}.
\end{eqnarray}
This quantity is related to the so called \textit{shear parameter}, which is bounded by the cosmic microwave background radiation and is a dimensionless measure of the anisotropy of the Universe (See chapter 5.2.2 of \cite{WE}).
Using this notation and with the help of (\ref{a}), (\ref{b}) and (\ref{yes}) we have
\begin{eqnarray}\label{no}
 \dot{F}=H[F(1-\frac{3}{2}F-\frac{8\pi }{H^2} \tr S)-\frac{16\pi}{H^3} S_{ab}\sigma^{ab}]
\end{eqnarray}
From the constraint equation (\ref{c})
\begin{eqnarray}\label{F}
 F=\frac{2}{3}-16\pi\frac{\rho}{H^2}
\end{eqnarray}
In general we see that $0\leq F\leq\frac{2}{3}$.\\

\subsection{Special cases}
\subsubsection{Vacuum}
We see that in the vacuum case $F=\frac{2}{3}$ and $\dot{F}=0$. Actually in this case one can write explicitly the solution known as \textit{Kasner solution}:
\begin{eqnarray*}
 ^{4}g=-dt^2+t^{2p_1}dx^2+t^{2p_2}dy^2+t^{2p_3}dz^2
\end{eqnarray*}
where the constants $p_i$ are called the \textit{Kasner exponents} which satisfy the two \textit{Kasner relations} given by:
\begin{eqnarray*}
 p_1+p_2+p_3&=&1 \\
(p_1)^2+(p_2)^2+(p_3)^2&=&1
\end{eqnarray*}
The mean curvature in this case is
\begin{eqnarray*}
 H=-t^{-1}
\end{eqnarray*}
Let $\lambda_i$ be the eigenvalues of $k_{ij}$ with respect to $g_{ij}$, i.e., the solutions of
\begin{eqnarray}\label{ev}
\det (k^i_j-\lambda \delta^i_j)=0 
\end{eqnarray}
 We define
\begin{eqnarray}\label{gke}
p_i=\frac{\lambda_i}{H} 
\end{eqnarray}
as the \textit{generalized Kasner exponents}. They satisfy the first but not in general not the second Kasner relation.

\subsubsection{Flat Friedmann with dust}
A special case of Bianchi I is the Friedmann model with flat spatial geometry. The metric in this case is:
\begin{eqnarray*}
^{4}g=-dt^2+a^2(dx^2+dy^2+dz^2)
\end{eqnarray*}
where $a$ is the scale factor. In this case $\sigma_{ab}=0$ which means that $F=0$ and also that $\dot{F}=0$.\\ Moreover in the dust case $S_{ab}=0$ (the Einstein-dust system can be thought of as singular case of the Einstein-Vlasov system, see chapter 3.4 of \cite{RA} for more information) and we can solve (\ref{yes}) obtaining:
\begin{eqnarray*}
 H=-2t^{-1}
\end{eqnarray*}
We can also write down the explicit solution of the metric in this case:
\begin{eqnarray*}
^{4}g=-dt^2+t^{\frac{4}{3}}(dx^2+dy^2+dz^2).
\end{eqnarray*}
This is also called the Einstein-de Sitter model.

\subsubsection{Bianchi I: the dust case (with small data)}
Let us look at the dust case not necessarily isotropic. The general solution is known and one can see from the solution (11-1.12) of \cite{HS} that the spacetime will isotropize. Nevertheless we will analyze this case with care, since we will show that the general case behaves asymptotically like it assuming small data. In the dust case:
\begin{eqnarray}\label{dd}
 \dot{F}=HF(1-\frac{3}{2}F)\leq 0
\end{eqnarray}
We are interested in the asymptotic behaviour at late times and will assume that $F$ is small, i.e.: $F < \epsilon$. Then it follows from (\ref{yes}):
\begin{eqnarray}\label{hd}
\dot{H}=(\frac{1}{2}+\frac{3}{4}F)H^2\leq(\frac{1}{2}+\epsilon)H^2
\end{eqnarray}
Integration leads to
\begin{eqnarray*}
 H \leq -\frac{2-\epsilon}{t+C}
\end{eqnarray*}
Using this inequality in (\ref{dd}) and integrating:
\begin{eqnarray}\label{YOY}
 F=O(t^{-2+\epsilon}).
\end{eqnarray}
We can put the equality (\ref{hd}) in the following form:
\begin{eqnarray}\label{II}
-H^{-1}=-H(t_0)^{-1}+\frac{1}{2}(t-t_0)+I_D
\end{eqnarray}
where $I_D$ is
\begin{eqnarray}
 I_D=\int^t_{t_{0}} \frac{3}{4} F(s)ds
\end{eqnarray}
Now we will use the fact that we can choose freely the time origin setting $t_{0}=-2H(t_0)^{-1}$. Note that for the general Bianchi I symmetric Einstein-Vlasov-system we know that $H$ takes all values in the range $(-\infty,0)$ (Lemma 2.1 of \cite{IS}) We then obtain
\begin{eqnarray}
 H+2t^{-1}\leq4 I_D t^{-2}
\end{eqnarray}
Using (\ref{YOY}) it is clear that $I_D\leq C(t_0)$. Note that with our time origin choice $C=0$ in (\ref{if}), so
\begin{eqnarray}\label{hi}
H\geq-2t^{-1} 
\end{eqnarray}
Our result for $H$ is the following:
\begin{eqnarray}\label{1}
 H=-2t^{-1}[1+O(t^{-1})]
\end{eqnarray}
This equation in (\ref{dd}) leads after integration to:
\begin{eqnarray}\label{2}
 F=O(t^{-2})
\end{eqnarray}
From (\ref{hi}) and (\ref{2}) we can conclude that:
\begin{eqnarray}\label{3}
 \sigma_{ab}\sigma^{ab}=O(t^{-4})
\end{eqnarray}

Now from (\ref{TF}) we see that the eigenvalues (\ref{ev}) of the second fundamental form with respect to the induced metric are also the solutions of
\begin{eqnarray*}
 \det(\sigma^i_j-[\lambda-\frac{1}{3}H]\delta^i_j)
\end{eqnarray*}
Let us define the eigenvalues of $\sigma_{ij}$ with respect to $g_{ij}$ by $\widehat{\lambda}_i$, we have that:
\begin{eqnarray*}
\widehat{\lambda}_i= \lambda_i-\frac{1}{3}H
\end{eqnarray*}
Note that $\Sigma_i (\widehat{\lambda}_i)^2=\sigma_{ab}\sigma^{ab}$. From (\ref{hi}) and (\ref{3}) we can see that the spacetime isotropizes at late times, in the sense that
\begin{eqnarray*}
 p_i= \frac{1}{3} +O(t^{-1})
\end{eqnarray*}
where $p_i$ are the generalized Kasner exponents.\\
Now using (\ref{det}) and (\ref{hi}) we have:
\begin{eqnarray*}
 \det g=O(t^4)
\end{eqnarray*}
So the hope would be to show that $\vert g_{ab} \ t^{-\frac{4}{3}} \vert$ can be bounded by a constant.
Let us define
\begin{eqnarray*}
\bar{g}_{ab}&=&t^{-\frac{4}{3}}g_{ab}\\
\bar{g}^{ab}&=&t^{+\frac{4}{3}}g^{ab}.
\end{eqnarray*}
 We have then with (\ref{a}) that:
\begin{eqnarray*}
 \dot{\bar{g}}_{ab}=-\frac{2}{3}(2t^{-1}+H)\bar{g}_{ab}-2t^{-\frac{4}{3}}\sigma_{ab}
\end{eqnarray*}
Making similar computations as in \cite{Lee} we arrive at:
\begin{eqnarray*}
\Vert \bar{g}_{ab}(t) \Vert \leq \Vert \bar{g}_{ab}(t_0) \Vert + \int^t_{t_0} [\frac{2}{3} \vert 2s^{-1}+H(s)\vert +2(\sigma_{ab}\sigma^{ab}(s))^{\frac{1}{2}}] \Vert \bar{g}_{ab}(s) \Vert ds
\end{eqnarray*}
and with Gronwall's inequality we obtain:

\begin{eqnarray}\label{tt}
\Vert \bar{g}_{ab}(t) \Vert \leq \Vert \bar{g}_{ab}(t_0) \Vert \exp\{ \int^t_{t_0} [\frac{2}{3} \vert 2s^{-1}+H(s)\vert +2(\sigma_{ab}\sigma^{ab}(s))^{\frac{1}{2}}]ds\}\leq C
\end{eqnarray}
Therefore $\bar{g}_{ab}$ is bounded for all $t\geq t_0$. The same holds for $\bar{g}^{ab}$ by similar computations. Thus:
\begin{eqnarray*}
\vert t^{-\frac{4}{3}}g_{ab} \vert \leq C\\
\vert t^{+\frac{4}{3}} g^{ab} \vert \leq C
\end{eqnarray*}
 From this we can conclude that:
\begin{eqnarray*}
 \Vert \sigma_{ab} \Vert \leq C t^{-\frac{2}{3}} 
\end{eqnarray*}
or
\begin{eqnarray*}
 \sigma_{ab} = O(t^{-\frac{2}{3}})
\end{eqnarray*}
Looking again at the derivative of $\bar{g}_{ab}$ and putting the facts which have been obtained together, we see that:
\begin{eqnarray*}
\dot{\bar{g}}_{ab}=O(t^{-2}) 
\end{eqnarray*}
This is enough to conclude that:
\begin{eqnarray*}
 g_{ab}=t^{+\frac{4}{3}}[\mathcal{G}_{ab}+O(t^{-2})]\\
g^{ab}=t^{-\frac{4}{3}}[\mathcal{G}^{ab}+O(t^{-2})]
\end{eqnarray*}
where $\mathcal{G}_{ab}$ is the limit of $\bar{g}_{ab}$ as $t$ goes to infinity.

\section{Bianchi I: the general case}

For the general case the basic equations are (\ref{yes}), a modified version of (\ref{b}) where $\gamma$ is a small and positive quantity which is introduced for technical reasons and (\ref{no})

\begin{eqnarray}
&&\partial_{t}(-H^{-1})=\frac{1}{2}+\frac{3}{4}F+4\pi \frac{\tr S}{H^2}\label{aa}
\\
&&\frac{d}{dt}(t^{-\gamma}\bar{g}^{ab})=t^{-\gamma}\bar{g}^{ab}[\frac{2}{3}(2t^{-1}+H)-\gamma t^{-1}]+2t^{-\gamma+\frac{4}{3}}\label{bb}
\sigma^{ab}
\\
&&\dot{F}=H[F(1-\frac{3}{2}F-\frac{8\pi }{H^2} \tr S)-\frac{16\pi}{H^3} S_{ab}\sigma^{ab}]\label{cc}
\end{eqnarray}

We have a number (different from zero) of particles at possibly different momenta and we will define $P$ as the supremum of the absolute value of these momenta at a given time $t$:
\begin{eqnarray}\label{P}
 P(t)=\sup \{ \vert p \vert =(g^{ab}p_a p_b)^\frac{1}{2} \vert f(t,p)\neq 0\}
\end{eqnarray}

\begin{theorem}
Consider any $C^{\infty}$ solution of the Einstein-Vlasov system with Bianchi I-symmetry and with $C^{\infty}$ initial data. Assume that $F(t_0)$ and $P(t_0)$ are sufficiently small. Then at late times one can make the following estimates:
\begin{eqnarray}
 H(t)&=&-2t^{-1}(1+O(t^{-1}))\\
P(t)&=&O(t^{-\frac{2}{3}+\epsilon})\label{Pp}\\
 F(t)&=&O(t^{-2})
\end{eqnarray}
\end{theorem}
\noindent
{\bf Remark} (\ref{Pp}) implies that asymptotically there is a dust-like behaviour (see (\ref{pp})). 

\noindent
{\bf Proof}
We will use a bootstrap argument (see chapter 10.3 of \cite{RA} for more information). Let us look at the interval $[t_0,t_1)$. Our bootstrap assumptions are the following:
\begin{eqnarray}
 &&F(t)\leq A(1+t)^{-\frac{3}{2}}\label{a1}\\
&&P(t)\leq B(1+t)^{-\frac{7}{12}}\label{a2}.
\end{eqnarray}
where A and B are positive constants which we can choose as small as we want.\\\\
\textit{1. Estimate of $H$}\\\\
Integrating (\ref{aa}) we obtain:
\begin{eqnarray}\label{H}
 -H^{-1}=-H(t_0)^{-1}+\frac{1}{2}(t-t_0)+I_G
\end{eqnarray}
with
\begin{eqnarray*}
 I_G=\int^t_{t_0}(\frac{3}{4}F+4\pi \frac{\tr S}{H^2})(s)ds\leq \int_{t_0}^t [\frac{3}{4} A (1+s)^{-\frac{3}{2}}+\frac{1}{4} \frac{\tr S}{\rho}(s)]ds
\end{eqnarray*}
where the inequality comes from (\ref{a1}) and (\ref{c}). Consider now an orthonormal frame and denote the components of the spatial part of the energy-momentum tensor in this frame by $\widehat{S}_{ab}$. The components can be bounded by
\begin{eqnarray}\label{pp}
\widehat{S}_{ab} \leq P^2(t) \rho
\end{eqnarray}
so we have that 
\begin{eqnarray}\label{ss}
\frac{\tr \widehat{S}}{\rho} \leq 3P^2 
\end{eqnarray}
from which follows that:
\begin{eqnarray*}
 I_G \leq \frac{18}{4}(A+B^2)
\end{eqnarray*}
Now returning back to (\ref{H}) and setting as in the dust case $t_0=-2H(t_0)^{-1}$ we have
besides
\begin{eqnarray}\label{ho}
 H\geq -2t^{-1}
\end{eqnarray}
that
\begin{eqnarray}\label{hh}
 H+2t^{-1}\leq 18(A+B^2)t^{-2}
\end{eqnarray}
\\\textit{2. Estimate of $P$}\\\\
Now we will use the second equation (\ref{bb}) to obtain an estimate for the metric. One can show that in the sense of quadratic forms the following is true:
\begin{eqnarray*}
\sigma^{ab} \leq (\sigma_{ab}\sigma^{ab})^{\frac{1}{2}} g^{ab}.
\end{eqnarray*}

Then with the definitions of $\bar{g}^{ab}$ and $F$ and the estimates (\ref{a1}) and (\ref{ho}) we have: :
\begin{eqnarray*}
 2t^{-\gamma+\frac{4}{3}}\sigma^{ab} \leq 2t^{-\gamma+\frac{4}{3}}(\sigma_{ab}\sigma^{ab})^{\frac{1}{2}} g^{ab}=2F^{\frac{1}{2}}\vert H \vert t^{-\gamma} \bar{g}^{ab}\leq 4 A^{\frac{1}{2}} (1+t)^{-\frac{3}{4}}t^{-1} t^{-\gamma} \bar{g}^{ab}
\end{eqnarray*}

Now from the last inequality of the first step (\ref{hh}) equation (\ref{bb}) leads to:
\begin{eqnarray*}
 \frac{d}{dt}(t^{-\gamma}\bar{g}^{ab})\leq t^{-\gamma}\bar{g}^{ab}[12(A+B^2)t_0 + 4 A^{\frac{1}{2}} (1+t_0)^{-\frac{3}{4}}-\gamma]t^{-1}=-\eta t^{-\gamma}\bar{g}^{ab}t^{-1}
\end{eqnarray*}
$A$ and $B$ have to be chosen in such a way that $\eta$ is positive, then
\begin{eqnarray}\label{neg}
\frac{d}{dt}(t^{-\gamma}\bar{g}^{ab})\leq 0
\end{eqnarray}
from which follows that
\begin{eqnarray*}
 \bar{g}^{ab}=O(t^{\gamma})
\end{eqnarray*}
or
\begin{eqnarray*}
 g^{ab}=O(t^{-\frac{4}{3}+\gamma})
\end{eqnarray*}
Now from (\ref{neg}) we have:
\begin{eqnarray*}
 t^{-\gamma+\frac{4}{3}}g^{ab}(t) \leq t_0^{-\gamma +\frac{4}{3}} g^{ab}(t_0)
\end{eqnarray*}
Using the fact that $p_a$ is constant along the geodesics we can conclude that:
\begin{eqnarray}\label{B}
 P(t)\leq {t_0}^{-\frac{\gamma}{2}+\frac{2}{3}}P(t_0)t^{\frac{\gamma}{2}-\frac{2}{3}}\leq Bt^{\frac{\gamma}{2}-\frac{2}{3}}
\end{eqnarray}
since we can choose $P(t_0)$ and $B$ independently as small as we want. In order to improve (\ref{a2}) $\gamma$ has to be smaller then $\frac{1}{6}$. Using the notation $\zeta=\frac{\gamma}{2}$ the last inequality can be expressed as
\begin{eqnarray}\label{b1}
 P(t)\leq B t^{-\frac{2}{3}+\zeta} 
\end{eqnarray}
where $\zeta < \frac{1}{12}$.\\\\
\textit{3. Estimate of $F$}\\\\
Until now we have an estimate for $H$ and for $P$ in the interval $[t_0,t_1)$. Now we have to improve the estimate for $F$ coming from the bootstrap assumption. The desired estimate is $F(t_1)\le A(1+t_1)^{-2+\epsilon}$. If this
is the case (case I) the bootstrap argument will work and there is nothing more to do.

\begin{center}{\includegraphics[height=6cm]{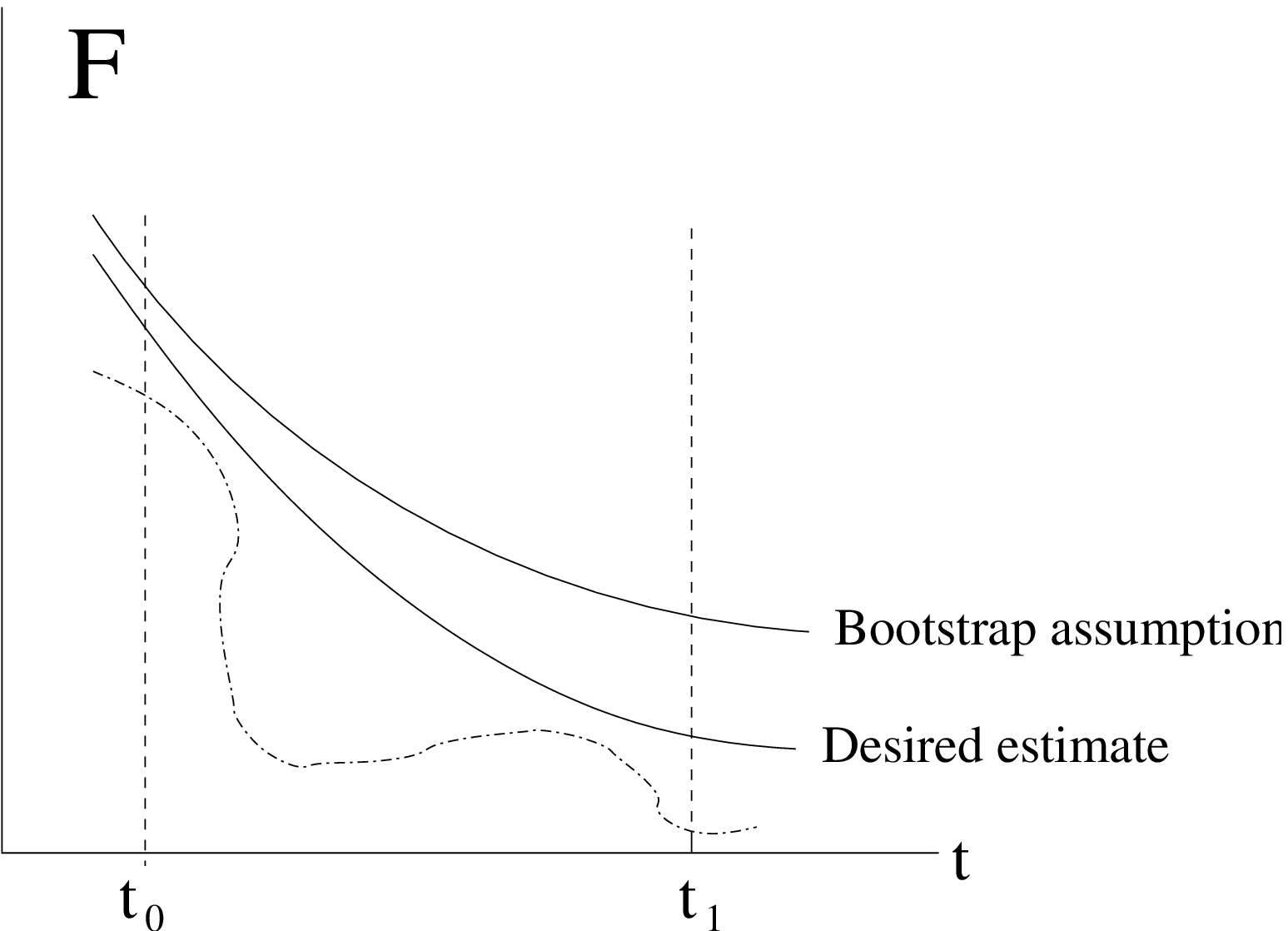}}
\textit{Case I}
\end{center}

Let us suppose now the opposite, that $F(t_1)> A(1+t_1)^{-2+\epsilon}$. Then define $t_2$ as the smallest number not smaller than $t_0$ with 
the property that $F(t_1)\geq A(1+t_1)^{-2+\epsilon}$. In this case, we have to distinguish between the case that $t_2 = t_0$ (Case IIa) 
and $t_2 > t_0$ (Case IIb). Let us look now at (\ref{cc}), in particular at the terms in square brackets. By using (\ref{a1})
\begin{eqnarray*}
 \frac{3}{2}F \leq \frac{3}{2} A (1+t)^{-\frac{3}{2}} \leq \frac{3}{2} A (1+t_0)^{-\frac{3}{2}} \leq \frac{\delta}{2}
\end{eqnarray*}
where $\delta$ is a positive small constant.
Using (\ref{c}), (\ref{a2}) and (\ref{ss}):
\begin{eqnarray*}
 8\pi \frac{ \tr \widehat{S}}{H^2} \leq \frac{3}{2}  P^2 \leq \frac{3}{2} B^2 (1+t)^{-\frac{7}{6}}\leq \frac{3}{2} B^2 (1+t_0)^{-\frac{7}{6}}\leq \frac{\delta}{4}
\end{eqnarray*}
With the Cauchy-Schwarz inequality, (\ref{c}), (\ref{pp}) and supposing that $F > A(1+t)^{-2+\epsilon}$ in the interval $[t_2,t_1]$:
\begin{eqnarray*}
 \vert 16 \pi \frac{\sigma_{ab}\widehat{S}^{ab}}{H^3} \vert \leq  F^{\frac{1}{2}}\frac{(\widehat{S}^{ab}\widehat{S}_{ab})^{\frac{1}{2}}}{\rho}\leq \sqrt{3}F F^{-\frac{1}{2}}P^2
 \leq\sqrt{3}  A^{-\frac{1}{2}}B^2 (1+t)^{-\frac{1}{6}-\frac{\epsilon}{2}}F \leq \frac{\delta}{4}F
\end{eqnarray*}
Note that although $A^{-\frac{1}{2}}$ may be a big quantity, since $A$ and $B$ are independent we can make $B$ smaller to ``correct'' this.
Using (\ref{hh}) and the last three inequalities in (\ref{cc}) leads to:
\begin{eqnarray*}
 \dot{F}\leq (-2+\varepsilon)(1-\delta)Ft^{-1}
\end{eqnarray*}
where $\varepsilon=18(A+B^2)t_0^{-1}$. Now setting $\xi=\varepsilon+2\delta-\varepsilon \delta$ we end up with:
\begin{eqnarray*}
 \dot{F} \leq (-2 +\xi )Ft^{-1}
\end{eqnarray*}
which means that
\begin{eqnarray}\label{b2}
 F(t_1)\leq F(t_2) {t_2}^{2-\xi} {t_1}^{-2+\xi} 
\end{eqnarray}

\textit{Case IIa.} In this case $t_2=t_0$, so (\ref{b2}) means
\begin{eqnarray*}
 F(t_1) \leq F(t_0) {t_0}^{2-\xi} {t_1}^{-2+\xi} \leq A {t_1}^{-2+\xi}
\end{eqnarray*}
since we can choose $F(t_0)$ as small as we want. So it follows that in this case:
\begin{eqnarray*}
 A {t_1}^{-2+\epsilon} \leq F(t_1) \leq  A {t_1}^{-2+\xi}.
\end{eqnarray*}
Situation which is schematically depicted in the following figure.

\begin{center}{\includegraphics[height=6cm]{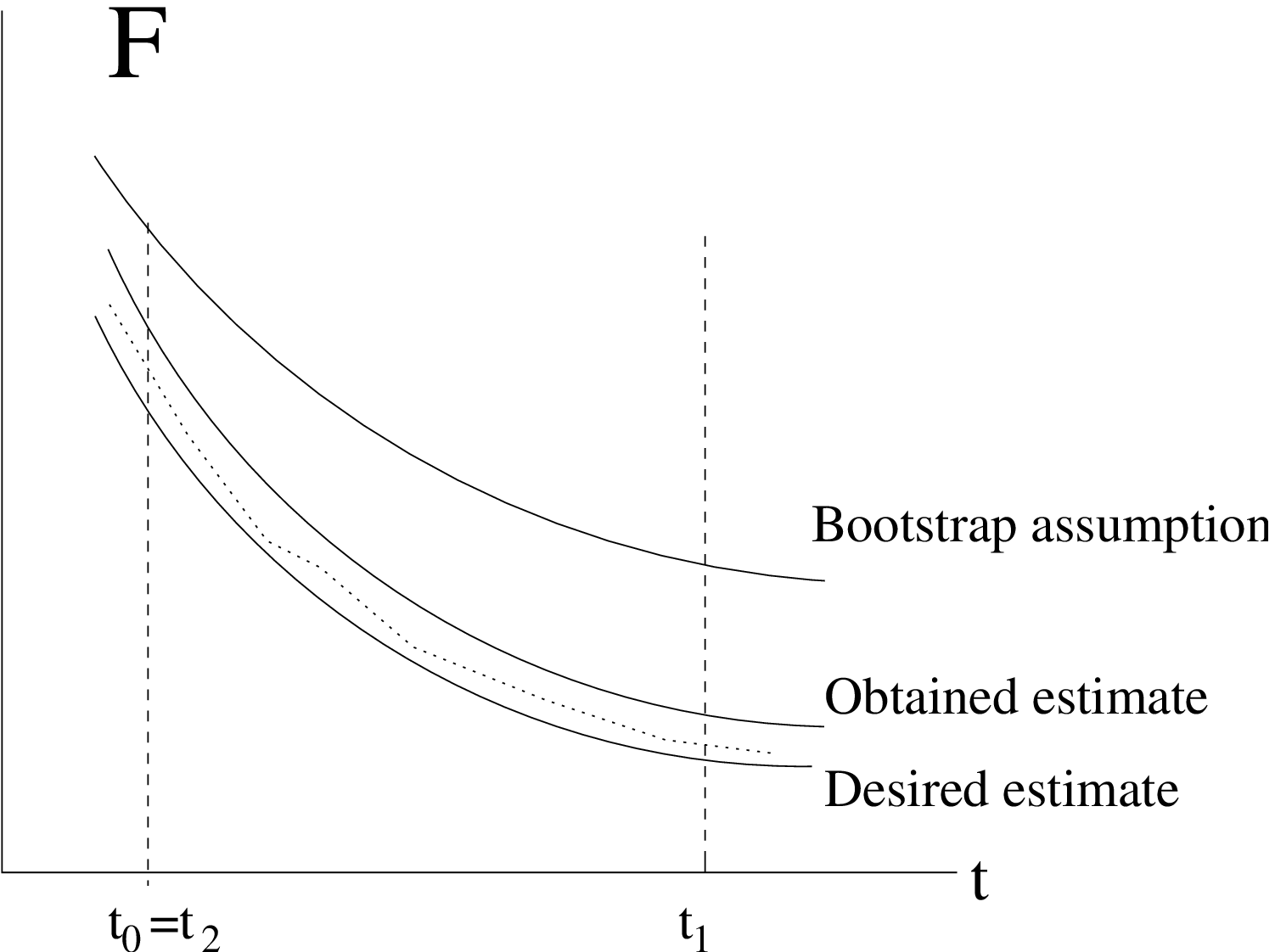}}
 \textit{Case IIa}
\end{center}

\textit{Case IIb.} In the case IIb we can use the fact that by continuity $F(t_2) \leq A (1+t_2)^{-2+\epsilon}$ holds and then
\begin{eqnarray*}
 F(t_1) \leq A (1+t_2)^{-2+\epsilon} {t_2}^{2-\xi} {t_1}^{-2+\xi} \leq A(1+t_2)^{\epsilon-\xi} {t_1}^{-2+\xi}
\end{eqnarray*}
The $\epsilon$ here is also a quantity which we can choose as small as we want and then it follows that in this case
\begin{eqnarray*}
 F(t_1) \leq A(1+t_0)^{\epsilon-\xi} {t_1}^{-2+\xi}
\end{eqnarray*}
We can choose $\epsilon$ to be smaller then $\xi$, so
\begin{eqnarray*}
 F(t_1) \leq A{t_1}^{-2+\xi}
\end{eqnarray*}

\begin{center}
\includegraphics[height=6cm]{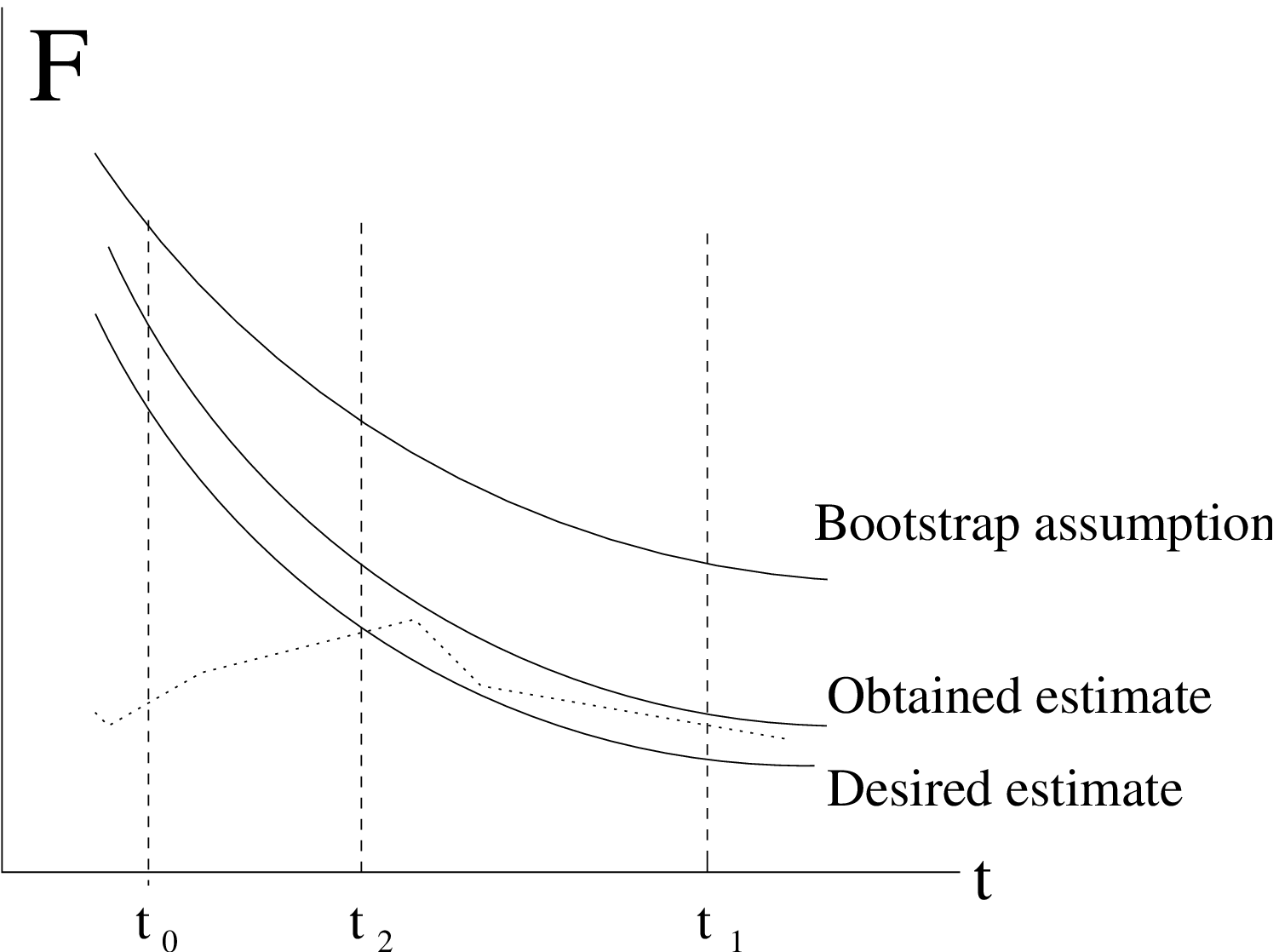}
\textit{Case IIb}
\end{center}

\textit{4. Results of the Bootstrap argument}\\

We have arrived at the statement that at least for a small interval $[t_0,t_1)$ and assuming (\ref{a1})-(\ref{a2}) we obtain the estimates:
\begin{eqnarray}
F(t)\leq A t^{-2+\xi}\label{yy}\\
 P(t)\leq B t^{-\frac{2}{3}+\zeta}\label{xx}
\end{eqnarray}
Thus both assumptions (\ref{a1}) and (\ref{a2}) have been improved and using a bootstrap argument we know that the estimates obtained are valid for $t_1=\infty$ assuming small data.
(\ref{hh}) can be expressed as:
\begin{eqnarray}\label{zz}
 H= -2t^{-1}(1+O(t^{-1}))
\end{eqnarray}
and is then also valid for the whole interval.\\\\

\textit{5. Improving the estimate of $F$}\\

We want to improve (\ref{yy}), but before that we need an inequality in the other
direction. From (\ref{cc}) we have:
\begin{eqnarray*}
 \dot{F} \geq H(F+\sqrt{3}F^{\frac{1}{2}}P^2)
\end{eqnarray*}
Implementing now (\ref{yy})-(\ref{xx}) in this inequality, in particular having in mind the 'B'-dependence of the estimate of $P$ (\ref{B}):
\begin{eqnarray*}
 \dot{F} \geq H F +H C(t_0)B^2 t^{-\frac{7}{3}+\epsilon}
\end{eqnarray*}
Using now (\ref{ho}):
\begin{eqnarray*}
 \dot{F} \geq -2t^{-1} F -C(t_0)B^2 t^{-\frac{10}{3}+\epsilon}
\end{eqnarray*}
from which follows
\begin{eqnarray*}
 t^2 F(t)\geq t_0^2 F(t_0)-\int^t_{t_0} C(t_0)B^2 s^{-\frac{4}{3}+\epsilon} ds
\end{eqnarray*}
and
\begin{eqnarray*}
 F(t)\geq t^{-2} (F(t_0)t_0^2-B^2 C(t_0){t_0}^{-\frac{1}{3}+\epsilon}+C(t_0)B^2t^{-\frac{1}{3}+\epsilon}) \geq t^{-2} (F(t_0)t_0^2-B^2 C(t_0){t_0}^{-\frac{1}{3}+\epsilon})
\end{eqnarray*}
$F(t_0)$ is strictly positive. Choosing now $B$ small enough we have the following estimate:
\begin{eqnarray}\label{U}
 F(t) \geq C(t_0)t^{-2}
\end{eqnarray}

Putting the last term of (\ref{cc}) with the help of (\ref{xx}) and (\ref{U}) in the following manner:
\begin{eqnarray*}
\vert 16 \pi \frac{\sigma_{ab}\widehat{S}^{ab}}{H^3} \vert \leq C F (F^{-\frac{1}{2}}P^2)=F O(t^{-\frac{1}{3}+\gamma})
\end{eqnarray*}
we can improve (\ref{yy}) implementing (\ref{yy})-(\ref{zz}) in (\ref{cc}) with the result:
\begin{eqnarray}\label{FF}
 F= O(t^{-2})
\end{eqnarray}
\\
With these results we obtain the following theorem.
\begin{theorem}\label{t2}
 Consider the same assumptions as in the previous theorem. Then
\begin{eqnarray}\label{ke}
 p_i= \frac{1}{3} +O(t^{-1})
\end{eqnarray}
and 
\begin{eqnarray}\label{lp}
g_{ab}=t^{+\frac{4}{3}}[\mathcal{G}_{ab}+O(t^{-2})]\\\label{ll}
g^{ab}=t^{-\frac{4}{3}}[\mathcal{G}^{ab}+O(t^{-2})] 
\end{eqnarray}
where $\mathcal{G}_{ab}$ and $\mathcal{G}^{ab}$ are independent of $t$.
\end{theorem}
\noindent
{\bf Proof} 
The conclusions made in the dust case after (\ref{2}) only depend on the estimate of $F$ and $H$ and apply directly to the general case. \\

\section{Conclusions and Outlook}
The result concerning the asymptotics of the metric implies ``asymptotic freezing'' in the expanding direction in the following sense. Consider the metric (4.12) of \cite{HN}, then by comparison with (\ref{lp}) one sees that the off-diagonal degrees of freedom $n_1$,$n_2$ and $n_3$ tend to constants and thus are not important for the dynamics. See \cite{HN} for the importance of asymptotic freezing in the ``other'' direction which has been studied, namely the initial singularity, and \cite{MK} for consequences of that in a quantum version.\\
As already mentioned in the introduction there exist several results concerning the Bianchi I-symmetric Einstein-Vlasov system (without cosmological constant). Almost all of them assume additional symmetries namely the LRS and the reflection symmetry condition (see \cite{IS} for a precise definition of these symmetries). However there exist some results where only the reflection symmetry is assumed. Concerning the expanding direction there is theorem 5.4 of \cite{IS}. Our theorems can be seen as a generalization of that theorem since we obtain the same the result, but a) we also obtain how fast the expressions converge b) we obtain an asymptotic expression for the spatial metric c) we do not assume any of the additional symmetries mentioned. However we used a different kind of restriction namely the small data assumptions.\\
In any case we think it interesting to study the non-diagonal case because although this time there was not an essential difference in the result with respect to the diagonal case, we do not expect that this will always be the case, especially when analyzing the initial singularity. In \cite{HU1} the possible dynamical behaviour towards the past has been determined assuming only the reflection symmetry and already there surprising new features like the existence of heteroclinic networks arose.\\
The situation of several (tilted) fluids leads naturally to the non-diagonal case. The Bianchi I-symmetry implies the absence of a matter current such that a single \textit{tilted} fluid is not compatible with this assumption. However there can be several tilted fluids such that the total current vanishes. In \cite{SU} and \cite{SA} this has been considered in the case of two fluids. What was found is that isotropization occurs if at least one of the two fluids has a speed of sound which is less or equal $\frac{1}{3}$ the speed of light. It is shown in particular for the case of two pressure free fluids in \cite{SA}, which can be seen as a singular solution of the Einstein-Vlasov system.\\
Of course there are many other ways of generalizing results which have been obtained for (single) perfect fluids. See for instance \cite{LB} and references therein for the inclusion of a Maxwell field in the Bianchi I case. In presence of a cosmological constant the results of \cite{Lee} have been generalized even to the Einstein-Vlasov-Maxwell case \cite{NT}. A natural generalization of the Vlasov equation is the case where the collision term is not zero, i.e. the Boltzmann equation. For this case dust-like asymptotics have already been obtained in \cite{ET} for an isotropic spacetime with a cosmological constant and \cite{ND} provides a basis for a possible extension to the asymptotics in the case of Bianchi I with LRS symmetry.\\
Finally we would like to mention that non-diagonal Bianchi I spacetimes are not only of interest in the context of cosmology, see for instance a recent work on the so called ultra-local limit \cite{CV}.

\textbf{Acknowledgements}\\
The author would like to thank Alan D. Rendall for helping along all the steps of this project, from proposing the problem to actually helping to solve it. I am also thankful for many discussions and a lot of concrete advice. This work has been funded by the Deutsche Forschungsgemeinschaft via the SFB 647-project B7.

\end{document}